\documentclass[pre,twocolumn,showpacs,preprintnumbers,amsmath,amssymb]{revtex4}

\usepackage{graphicx}
\usepackage{dcolumn}
\usepackage{bm}

\begin{document}


\title{Dynamic range of hypercubic  stochastic excitable media}

\author{Vladimir R. V. Assis}%
\email{vladimirassis@df.ufpe.br}
\author{Mauro Copelli}%
 \email{mcopelli@df.ufpe.br}
\thanks{corresponding author}
\affiliation{%
Laborat\'orio de F{\'\i}sica Te\'orica e Computacional, Departamento
de F{\'\i}sica, Universidade Federal de Pernambuco, 50670-901 Recife, PE, Brazil}%



\begin{abstract}
We study the response properties of $d$-dimensional hypercubic
excitable networks to a stochastic stimulus. Each site, modelled
either by a three-state stochastic
susceptible-infected-recovered-susceptible system or by the
probabilistic Greenberg-Hastings cellular automaton, is continuously
and independently stimulated by an external Poisson rate $h$.  The
response function (mean density of active sites $\rho$ versus $h$) is
obtained via simulations (for $d=1, 2, 3, 4$) and mean-field
approximations at the single-site and pair levels ($\forall \, \, d$). In
any dimension, the dynamic range and sensitivity of the response
function are maximized precisely at the nonequilibrium phase
transition to self-sustained activity, in agreement with a reasoning
recently proposed. Moreover, the maximum dynamic range attained at a
given dimension $d$ is a decreasing function of $d$.
\end{abstract}

\maketitle

\section{\label{intro} Introduction}

The building blocks of sensory organs are excitable neurons, upon
which physical stimuli impinge continuously. Information about these
stimuli is transformed into electrical activity of the neuronal
membranes, usually in the form of nonlinear excitations called
spikes. The biophysics of ion channels involved in this process has
been thoroughly investigated in the last five decades, after the
success of the Hodgkin-Huxley theory~\cite{HH52}. Despite immense
progress in this regard, however, some fundamental questions have
remained unanswered. One of them has to do with two apparently
conflicting experimental results. On the one hand, animals are
subjected to stimulus intensities spanning {\em many\/} orders of
magnitude, which their brains somehow manage to handle. This result is
perhaps most easily revealed in psychophysical experiments. When
humans are asked to assign an arbitrary (psychological) value to a
given physical stimulus, this value is shown to be proportional to a
power $m$ of the stimulus intensity~\cite{Stevens} (Stevens' law of
psychophysics). The fact that the Stevens' exponent $m$ is usually $<1$
is consistent with the large dynamic range and sensitivity of
psychophysical response functions (for instance, $m\simeq 0.6$ for the
smell of heptane~\cite{Stevens}). On the other hand, the response
(mean firing rate) of isolated sensory neurons as a function of
stimulus intensity has been shown to be an approximately linear
saturating curve, at least for some sensory modalities. This implies
that their dynamic range is usually small (for olfactory sensory
neurons, they stay in the range of $\sim 10$~dB~\cite{Rospars00,Rospars03}).

How is it then that large dynamic ranges are obtained from elements
which individually have small dynamic ranges? What is the mechanism
that generates Stevens' exponents $m<1$? Two main mechanisms have been
historically recognized as contributing to the phenomenon: one of them
invokes the intrinsic variation of thresholds in a population of
sensory neurons~\cite{Cleland99}. The second one is adaptation, by
which neurons adjust their operating ranges according to the
statistics of the ambient stimulus (see,
e.g., Refs.~\cite{Normann74,Werblin74a,Werblin74b,Kim03,Borst05} for
the case of the visual system). Both mechanisms certainly contribute
to an enhancement of dynamic range. However, recent experimental data
strongly suggest that additional mechanisms based on a collective
neuronal phenomenon could be at play: Deans et al.~\cite{Deans02}
showed that knocking out gap junctions (electrical synapses among
neurons) leads to a substantial change in the response function of
retinal ganglion cells of mammals, with a decrease in dynamic range
and sensitivity~\cite{Deans02} and an increase in the response
exponent~\cite{Furtado06}. 

In the last few years we have investigated this third possible
mechanism, which addresses how neurons could {\em cooperatively\/}
lead to an enhancement of dynamic range owing to the presence of
lateral interactions (e.g. via chemical or electrical
synapses). Connected, they form an extended system in which excitable
waves are created upon incidence of incoming stimuli and annihilated
upon collision (either with one another or with boundaries) due to the
nonlinearity of their dynamics. The overall effect of this process is
to collectively produce an enhancement of dynamic range and
sensitivity, as compared to those of the elements
alone~\cite{Copelli02, Copelli05a, Copelli05b, Furtado06, Kinouchi06a,
Wu07, Copelli07}. We emphasize that our proposal relies on very basic
properties of excitable media, which opens the possibility that it
could be applied not only to sensory systems, but wherever else
enhanced sensitivity and dynamic range are required. For example,
extreme sensitivity is observed in rat motor cortex, where stimulation
of a single pyramidal cell can evoke whisker
movements~\cite{Brecht04}. Also potentially related to what we propose
is the experimental observation that electrical synapses in the
neocortex are present exclusively between {\em inhibitory\/}
interneurons~\cite{Hestrin05}. Electrical coupling allows them to {\em
excite\/} each other~\cite{Hestrin05} and might have the functional
role of augmenting their dynamic range and sensitivity.

The mechanism we discuss here is simple: let the arrival of a
suprathreshold stimulus reaching a sensory neuron be modelled by a
Poisson process with rate $h$, which would be proportional to the
stimulus intensity (say, the concentration of an odorant reaching the
olfactory epithelium). For very small $h$, stimulus events are rare
and each of them would produce on average one excitation, if the
excitable elements were disconnected. If they are connected, however,
excitations can propagate stochastically to neighbors at some rate
$\lambda$. In this case, a single-stimulus event will generate an
amplifying excitable wave. If $\lambda$ is small, this wave will die
out after some time, but the average network activity $\rho$ will
nonetheless be larger than that of uncoupled neurons, leading to an
enhanced sensitivity and larger dynamic range. This amplifying effect
becomes more pronounced as $\lambda$ increases, so the dynamic range
initially increases with $\lambda$. Increasing $\lambda$ further,
however, one may reach a phase transition at some critical value
$\lambda=\lambda_c$, above which self-sustained activity becomes
stable [$\rho(h=0; \lambda>\lambda_c)>0$]. In this supercritical
regime, the larger the coupling $\lambda$, the more difficult it
becomes for $\rho$ to code for weak stimuli, which can hardly be
distinguished from the self-sustained background activity of the
network. Therefore, for $\lambda>\lambda_c$ the dynamic range
decreases with increasing $\lambda$. Putting those two results
together, one concludes that the dynamic range is maximum at
criticality~\cite{Kinouchi06a}.

Clearly, the above reasoning applies to essentially any network
topology. In its original version~\cite{Kinouchi06a}, it was
formulated for an Erd{\H o}s-R\'enyi random graph, where a simple mean-field model perfectly captured the phenomenon. However, the
enhancement of the dynamic range in that topology was about $50\%$,
which is much less than what is observed experimentally (for instance,
dynamic ranges in the olfactory glomerulus are at least twice as large
as in olfactory sensory neurons~\cite{Friedrich97,Wachowiak01}). This
raises the question whether networks with different topologies could
yield larger dynamic ranges and, if so, how these depend on network
structural parameters. In Ref.~\cite{Copelli07}, for instance, the
dynamic range of scale-free networks was shown to depend on the
density of loops, but the evidence relied only on numerical
simulations. In this contribution, we make use of simulations and
analytical methods to deal with this question in hypercubic lattices,
looking at the dimension $d$ as a parameter.

\section{\label{model} model and simulation results}

We explore these ideas within a simple stochastic model of
pulse-coupled excitable elements:

\begin{eqnarray}
\label{eq:p0}
\dot P_t(S_x) & = & - hP_t(S_x) - \lambda \sum_{y}P_t(S_x,I_y) + \gamma P_t(R_x)  \\
\label{eq:p1}
\dot P_t(I_x) & = & -P_t(I_x) + \lambda \sum_{y}
P_t(S_x,I_y) + hP_t(S_x)  \\
\label{eq:p2}
\dot P_t(R_x) & = & -\gamma P_t(R_x) + P_t(I_x)\; ,
\end{eqnarray}
where $P_t(\alpha_x)$ is the probability that site at location
$x\in\{1,\ldots,N\}$ is in state $\alpha$ at time $t$; $P_t(\alpha_x,
\beta_y)$ is the joint probability that sites at locations $x$ and $y$
are respectively in states $\alpha$ and $\beta$ at time $t$; $\alpha,
\beta \in \{S,I,R\}$ denote a quiescent, excited or refractory state,
respectively; $y$ runs over the neighborhood of $x$; and $\gamma^{-1}$
is the characteristic refractory time, measured in units of the
characteristic excitation time (defined as 1, without loss of
generality~\cite{Joo04b}). We employ the notation of the stochastic
susceptible-infected-recovered-susceptible (SIRS) model, to which this
model is identical except for the external-stimulus field $h$, which
is often missing in epidemiological modelling (where it amounts to a
spontaneous infection rate~\cite{Marro99}). We can therefore extend a
previous analysis of the stochastic SIRS model on a hypercubic lattice
by Joo and Lebowitz~\cite{Joo04b}, from which our results can be
derived if the external field $h$ is added as in
Eqs.~(\ref{eq:p0})-(\ref{eq:p2}).

\begin{figure}[!hbt]
\centerline{\includegraphics[width=0.95\columnwidth]{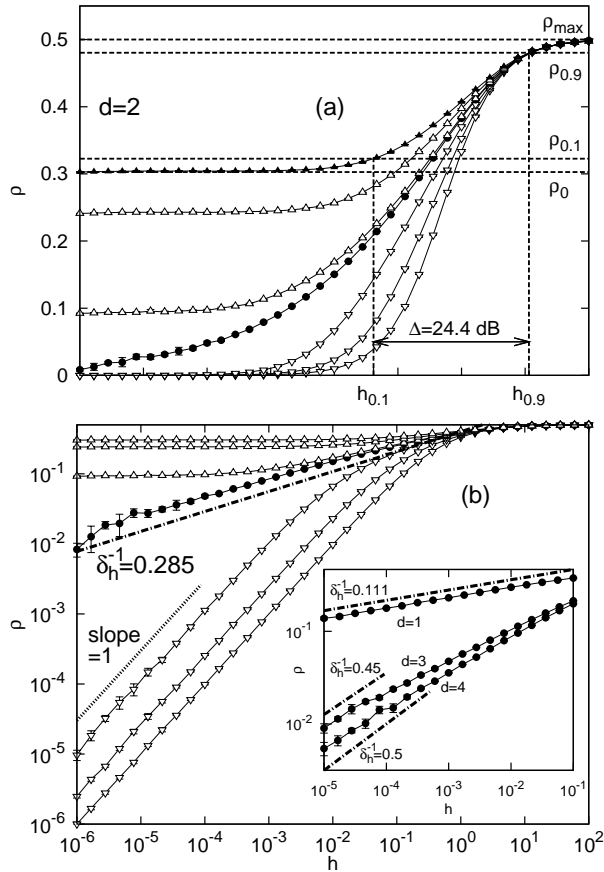}}
\caption{\label{fig:response}Response curves $\rho(h)$ of the
  stochastic SIRS model on a two-dimensional lattice (results from
  simulations with $N=100^2$ sites, periodic boundary conditions,
  averaged over a maximum time $T_{max} \sim 10^4-10^{6}$ and five
  runs) in (a) linear-log scale and (b) log-log scale. From bottom to
  top, triangles denote $\lambda=0,0.2,\ldots,1$. Filled circles
  denote $\lambda\simeq \lambda_c$. Relevant parameters for
  calculating the dynamic range $\Delta$ are exemplified in (a) for
  $\lambda=1$ (filled triangles). (b) Dot-dashed lines show literature
  values for the critical exponent $\delta_h^{-1}$. Inset: response
  function near criticality for $d=1,3$~and~4 (system sizes and
  approximate critical coupling are $N=5000$ and $\lambda_c\simeq
  7.73(8)$, $N=20^3$ and $\lambda_c\simeq 0.259(3)$, $N=10^4$ and
  $\lambda_c\simeq 0.167(2)$, respectively).}
\end{figure}

We are interested in the response function (or transfer function) of
the excitable medium---i.e. the dependence of the stationary density of
active sites $\rho \equiv\lim_{t\to\infty} P_t(I)$ on the external
stimulus intensity $h$ (note that, in the context of neuroscience, the
mean firing rate can be obtained by dividing $\rho$ by the mean
excitation time). Figure~\ref{fig:response} shows simulation results
which confirm the general scenario described above. For $\gamma=1$
(which is kept fixed throughout this paper) and $d=2$, a phase
transition occurs at $\lambda = \lambda_c \simeq 0.567(2)$. In the
subcritical regime ($\lambda<\lambda_c$) $\rho(h)$ is a
linear-saturating curve whose slope increases monotonically with
$\lambda$. In the supercritical regime ($\lambda>\lambda_c$) one has
$\rho_0 \equiv \lim_{h\to 0} \rho(h) > 0$, with $\rho_0$ increasing
monotonically with $\lambda$~\cite{Joo04b}.

\begin{figure*}
\includegraphics[width=0.85\textwidth]{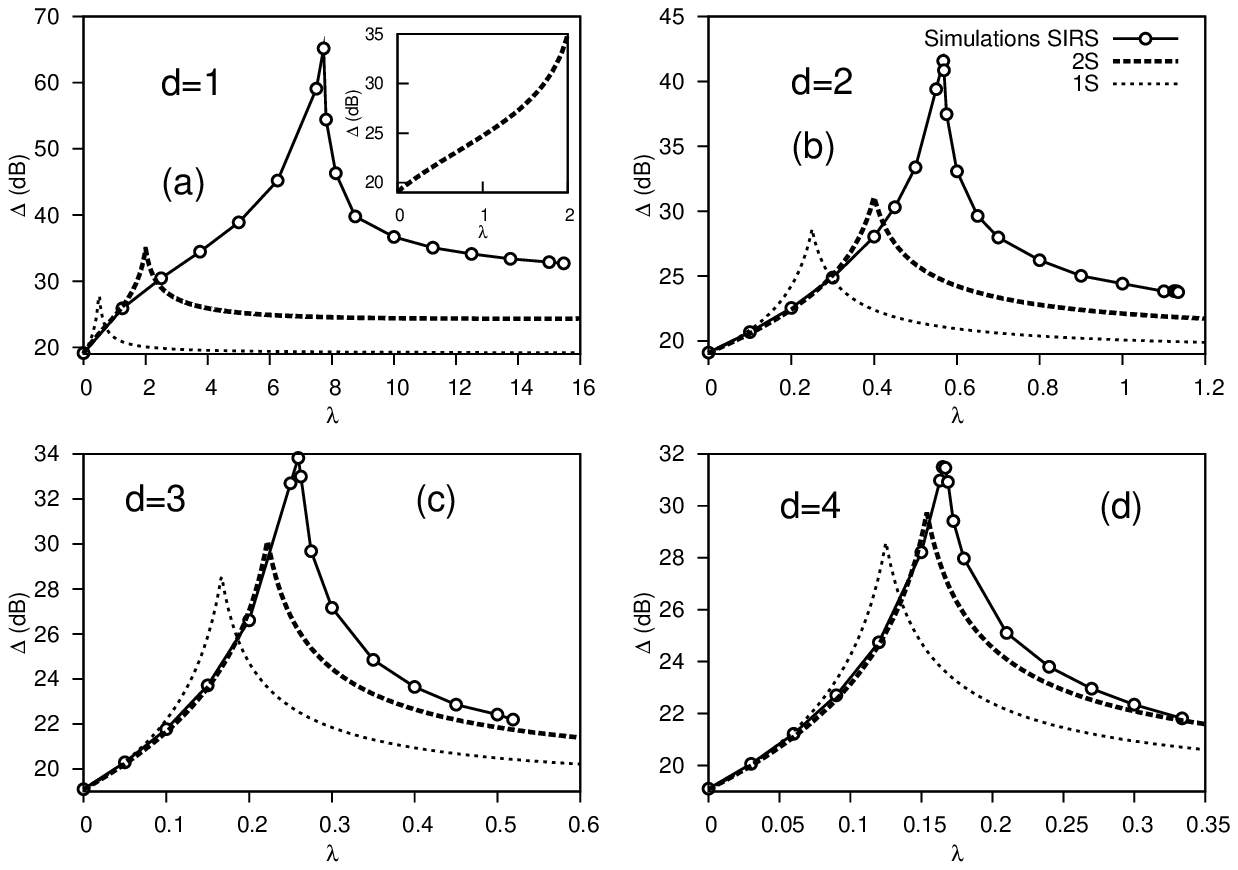}
\caption{\label{fig:dynrange}Dynamic range vs. coupling: simulation
  results (symbols), 1S (thin dashed) and 2S (thick dashed) mean field
  approximations. The peaks occur always at the phase
  transition. Half-widths of tuning of the critical regime [with
  heights measured relative to $\Delta(\lambda=0)$] are
  $\Delta\lambda\simeq 1.4$, $0.094$, $0.059$ and $0.042$ for
  $d=1,2,3$~and~4, respectively.}
\end{figure*}

To obtain the dynamic range of the response curve, we employ the
definition usually adopted in the biological
literature~\cite{Rospars00,Rospars03}. Let $h_{0.1}$ and $h_{0.9}$ be
the stimulus intensities such that their corresponding responses
($\rho_{0.1}$ and $\rho_{0.9}$) are, respectively, $10\%$ and $90\%$ above the
base-line activity within the range $[\rho_0,\rho_{max}]$:
\begin{equation}
\label{eq:rhox}
\rho_{\eta} \equiv \rho_0 + \eta(\rho_{max} - \rho_0)\; ,
\end{equation}
where $\rho(h_\eta)\equiv \rho_\eta$, $\rho_{max}\equiv
\lim_{h\to\infty}\rho(h) = \gamma/(\gamma+1)$, and $\eta \in
\{0.1,0.9\}$. The dynamic range $\Delta$ is defined as

\begin{equation}
\label{eq:delta}
\Delta \equiv 10\log_{10}\left(\frac{h_{0.9}}{h_{0.1}}\right)\; .
\end{equation}
As depicted in Fig.~\ref{fig:response}(a), $\Delta$ amounts to the
range of stimulus intensities (measured in dB) that can be
appropriately coded by the average activity in the network, discarding
stimuli whose responses are too close either to baseline activity
($\rho < \rho_{0.1}$) or to saturation ($\rho > \rho_{0.9}$).

\begin{figure}[t]
\centerline{\includegraphics[width=0.95\columnwidth]{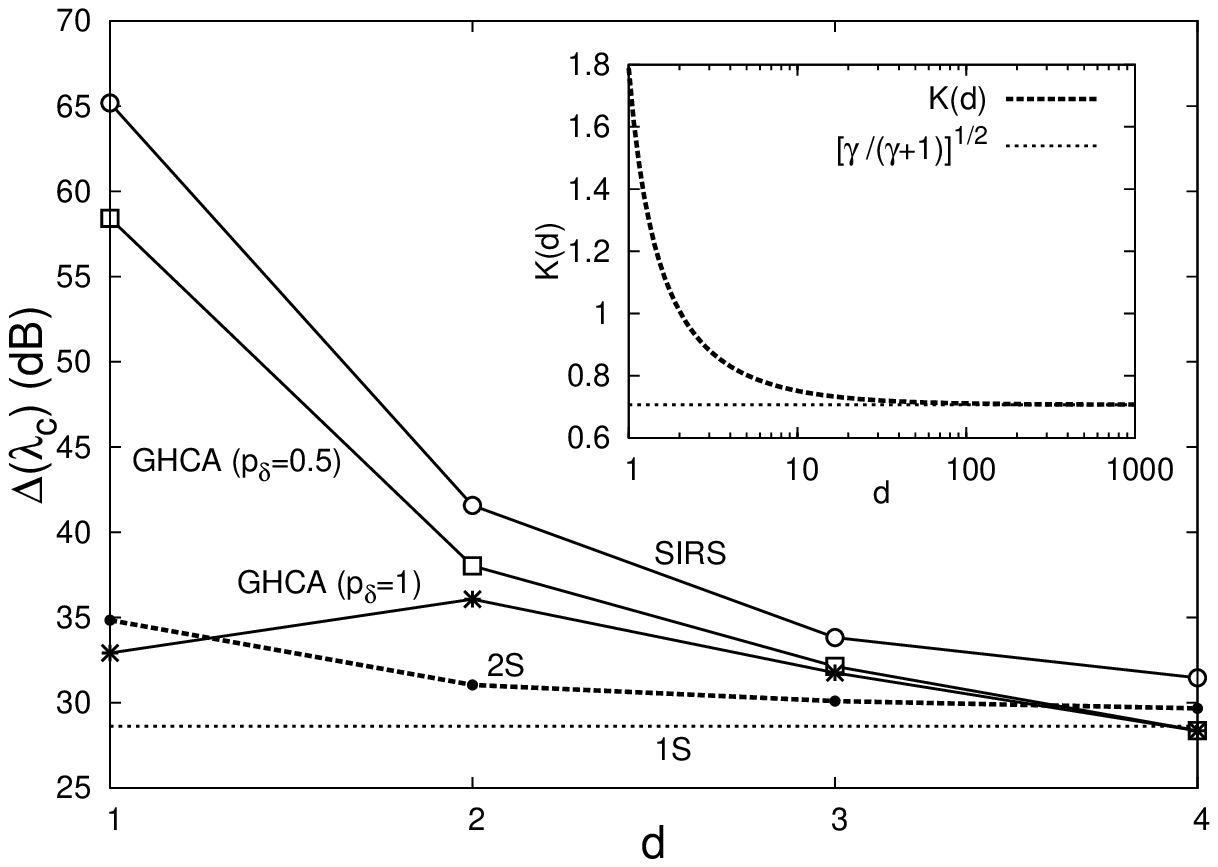}}
\caption{\label{fig:picos}Maximum dynamic range vs. dimension of the
  lattice: simulation results for the SIRS (open circles) and GHCA
  (stars denote $p_\delta=1$ and squares denote $p_\delta=0.5$)
  models, 1S (thin dashed) and 2S (thick dashed) approximations. GHCA
  simulations were performed with $p_\gamma=1/3$ and the same number
  of sites, $T_{max}$ and number of runs as described in the caption
  of Fig.~\ref{fig:response} for the SIRS model.  Inset: $K(d)$
  approaches the 1S coefficient $\rho_{max}^{1/2}$ in the limit
  $d\to\infty$ (see eq.~\ref{eq:K}).}
\end{figure}

According to this standard definition, the dynamic range of the
response curves in Fig.~\ref{fig:response}(a) shows the predicted
behavior: for $\lambda<\lambda_c$ ($\lambda>\lambda_c$),
$\Delta(\lambda)$ is a monotonically increasing (decreasing)
function. As shown in Fig.~\ref{fig:dynrange}, the maximum dynamic
range occurs precisely at criticality, where the response function is
governed by the scaling relation $\rho \sim h^{\delta_h^{-1}}$. In all
our simulations, the observed critical exponent $\delta_h^{-1}$ is
compatible with the literature values for the directed percolation
(DP) universality class: namely, $\delta_h^{-1}=0.111$, $0.285$,
$0.45$, and $1/2$ for $d=1$, $2$, $3$, and $d\geq 4$,
respectively~\cite{Adler&Duarte1987,Adler&Duarte1988,Obukhov1980} [see
inset of Fig.~\ref{fig:response}(b)]. This should be expected, since
the model has local rules, a continuous transition to a unique
absorbing state, and no further conservation
laws~\cite{Marro99,Janssen81,Janssen85,Grassberger82}. Because the
exponent $\delta_h^{-1}$ increases with increasing $d$ and since it
is the key element governing the dynamic range at criticality, we come
to one of the main results of this paper: the maximum dynamic range
attained at a given dimension $d$ is a decreasing function of
$d$. This result is summarized in Fig.~\ref{fig:picos}, which exhibits
the peaks of Fig.~\ref{fig:dynrange} versus the dimension of the lattice.

We further note that similar results are obtained if one employs the
Greenberg-Hastings cellular automaton (GHCA), where now the state
transition $S\to I$ is controlled by probabilities $p_h$ (external
stimulus) and $p_\lambda$ (coupling), $I\to R$ by $p_\delta$ and $R\to
S$ by $p_\gamma$~\cite{Greenberg78}. The only difference occurs for
the particular case of excitations with a deterministic duration
($p_\delta=1$) in $d=1$, in which case self-sustained activity cannot
be stable and the maximum dynamic range occurs for $p_\lambda=1$ with
an anomalous response exponent $1/2$, as previously
reported~\cite{Furtado06} (see stars in Fig.~\ref{fig:picos}). Note
that $p_\delta=1$ seems biologically more realistic for neurons: while
coupling may be stochastic, the duration of a spike is generally well
described by a deterministic dynamics. For $p_\delta<1$ and $d=1$ a
phase transition can occur just like in the SIRS model, and in this
case $\Delta(\lambda_c)$ is again a monotonically decreasing function
of $d$ (as exemplified for $p_\delta=0.5$; see squares in
Fig.~\ref{fig:picos}).

\section{\label{mf} Mean-field results}

\subsection{Single-site approximation}

The remainder of this paper focuses on the possibilities of
understanding these results with analytical means. Clearly, solving
Eqs.~(\ref{eq:p0})-(\ref{eq:p2}) for $\lambda\neq 0$ is difficult, because
the dynamics of single-site probabilities $P_t(\alpha_x)$ depend on
two-site joint probabilities $P_t(\alpha_x,\beta_y)$, which in their
turn depend on higher-order terms, and so forth. The simplest way to
truncate this infinite hierarchy of equations is the single-site (1S)
mean field approximation, in which correlations are
neglected~\cite{Marro99,Joo04b}. The conditional probabilities are
approximated as $P_t(\alpha_x | \beta_y)\stackrel{1S}{\simeq}
P_t(\alpha_x)$, which implies that $m$-site joint probabilities get
factorized as $P(\alpha^{(1)}_{x_1},\ldots,\alpha^{(m)}_{x_m})
\stackrel{1S}{\simeq} \prod_{j=1}^m P_t(\alpha^{(j)}_{x_j})$, where
$\alpha^{(j)} \in \{S,I,R\}$. In this approximation, assuming
homogeneity and isotropy, Eqs.~(\ref{eq:p0})-(\ref{eq:p2}) reduce to a
closed system and one obtains
\begin{eqnarray}
\rho(h) & \stackrel{1S}{=} & 
\left(\frac{\rho_{max}}{2\sigma}\right)(\rho^{-1}_{max}h+1-\sigma) \times
\nonumber \\
& & \left\{-1 + \sqrt{1 + \frac{4\sigma
\rho^{-1}_{max}h}{(\rho^{-1}_{max}h+1-\sigma)^2} }\right\}\; .
\label{eq:mf}
\end{eqnarray}
Note that information about the dimension $d$ of the network is
absorbed into an effective branching parameter $\sigma\equiv \lambda
z$, where $z\equiv 2d$ is the number of neighbors each site has. When
$\sigma\to 0$, we obtain a linear saturating response $\rho(h) =
\rho_{max}h/(\rho_{max}+h)$, which is an exact result for uncoupled
excitable elements. If $h\to 0$, Joo and Lebowitz's 1S results are
recovered: without an external stimulus, the 1S approximation predicts
a phase transition at $\sigma = \sigma_c = 1$ ($\lambda = \lambda_c =
1/z$), above which an active phase with $\rho_0 =
\rho_{max}\sigma(\sigma-1) > 0$ is stable~\cite{Joo04b}. The weak
stimulus response is linear in the subcritical regime,
$\rho(h;\lambda<\lambda_c) \stackrel{1S}{\simeq}
h/(1-z\lambda)$. At criticality, however, it is governed by
$\rho(h;\lambda=\lambda_c) \stackrel{1S}{\simeq} \left(\rho_{max}
h\right)^{1/2}$, which leads to the mean-field exponent
$\delta_h^{-1}=1/2$. Applying the definition of
Eqs.~(\ref{eq:rhox})~and~(\ref{eq:delta}) to Eq.~(\ref{eq:mf}), one obtains
the dynamic range in the 1S approximation:
\begin{equation}
\Delta(\sigma) \stackrel{1S}{=} \left\{            
\begin{array}{ll}
10\log_{10}\left[81\left(\frac{1-0.1\sigma}{1-0.9\sigma}\right)\right]&(\sigma\leq 1) \\
10\log_{10}\left[81\left(\frac{\sigma-0.1}{\sigma-0.9}\right)\right]&(\sigma\geq
1)\; . \\
\end{array}
\right.
\end{equation}
As depicted in Fig.~\ref{fig:dynrange}, the peak at $\sigma=1$ is
$\Delta(\sigma_c)=30\log_{10}9$~dB, which corresponds to an exact
$50\%$ enhancement as compared to uncoupled elements:
$\Delta(0)=20\log_{10}9$~dB.  Not surprisingly, this result holds
for all $d$.

\subsection{\label{pa} Pair approximation}

To gain analytical insight into the dependence of $\Delta$ on $d$, we
have solved Eqs.~(\ref{eq:p0})-(\ref{eq:p2}) in the so-called pair or
two-site (2S) approximation, in which conditional probabilities are
truncated beyond nearest neighbors: $P(\alpha_{x_1} | \beta_{x_2},
\chi_{x_3}) \stackrel{2S}{\simeq} P(\alpha_{x_1} | \beta_{x_2})$ (this
approximation is valid for the hypercubic lattice because $x_3$ is not
a nearest neighbor of $x_1$). This leads to three-site joint
probabilities being approximated as $P(\alpha_{x_1}, \beta_{x_2},
\chi_{x_3}) \stackrel{2S}{\simeq} P(\alpha_{x_1},\beta_{x_2})
P(\beta_{x_2},\chi_{x_3})/P(\beta_{x_2})$. To arrive at a closed set
of equations, first one has to write down the dynamics also for
two-site probabilities, which in this case are a direct extension of
Joo and Lebowitz's equations for $h\neq 0$~\cite{Joo04b}. These
equations
(\ref{evolucao_temporal_dos_pares_1})-(\ref{evolucao_temporal_dos_pares_3})
can be found in the Appendix. Applying the 2S approximation to all
equations, one concludes, after some manipulation, that $\rho$
satisfies a cubic equation [Eq.~(\ref{eq:cubic})]. For $h=0$ one can
obtain the 2S value of the coupling at which the phase transition
occurs: $\lambda_c \stackrel{2S}{=}
(\gamma+1)/\left(2d-2+(2d-1)\gamma\right)$~\cite{Joo04b}.

Figure~\ref{fig:dynrange} shows the dynamic ranges calculated from the
numerically obtained response curves. The 2S approximation shows a
better agreement with simulations than does the 1S approximation, and
Fig.~\ref{fig:dynrange} clearly shows how the agreement improves with
increasing $d$. In particular, note that the 2S approximation
reproduces the inflection in the $\Delta(\lambda)$ curve, which
appears only for $d=1$ [inset of Fig.~\ref{fig:dynrange}(a)]. Like in
the 1S mean-field approximation and in the simulations, the
subcritical response function for weak stimuli is linear:
\begin{equation}
\rho(h; \lambda<\lambda_c) \stackrel{2S}{\simeq} \left\{
\frac{\gamma+1 +
\lambda(\gamma+2)}{\left[2d-2+(2d-1)\gamma\right](\lambda_c -
\lambda)} \right\} h\; .
\end{equation}
More important for the topic of this article, the 2S approximation
manages to capture the dependence of $\Delta(\lambda_c)$ on $d$,
despite the fact that the weak-stimulus response at the critical
coupling is still governed by a mean-field exponent: $\rho(\lambda_c;
d) \stackrel{2S}{\simeq} K(d)h^{1/2}$. In this case, the decreasing
function
\begin{equation}
\label{eq:K}
K(d) = 
\sqrt{\frac {4 d^2 \gamma (\gamma + 1)^2 (2 d - 1)^{-1} } {\gamma^3
(2 d - 1) + \gamma^2 (6 d - 4) + \gamma (6 d - 5) + 2 d - 1}} 
\end{equation}
correctly incorporates the influence of $d$ on $\Delta(\lambda_c)$
(see inset of Fig.~\ref{fig:picos}).

\begin{figure}[t]
\centerline{\includegraphics[width=0.99\columnwidth]{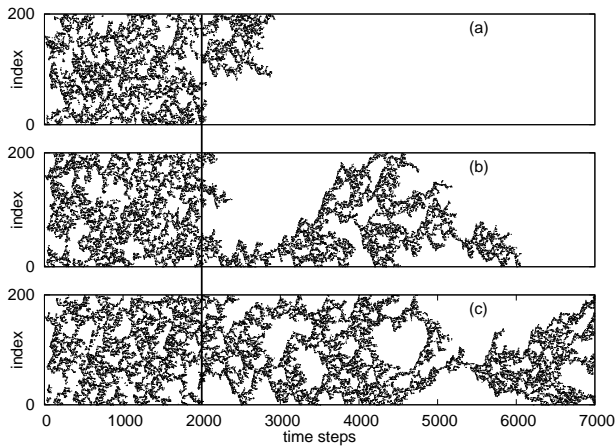}}
\caption{\label{fig:singlerun}Active sites vs. time in a single run of
  a small $d=1$ SIRS system (one time step is counted every $N=200$
  updates; boundary conditions are open). An external stimulus with
  rate $h=10^{-2}$ is applied for $0<t<2000$ ($h=0$ otherwise). Upper,
  middle and lower panels show the subcritical ($\lambda=7.0$),
  critical ($\lambda=7.73$) and supercritical ($\lambda=8.5$) regimes,
  respectively. }
\end{figure}

The fact that neither the 1S nor 2S approximations can reproduce
the correct critical exponents for $d<4$ is well known. Inherent to
those mean-field approximations is the truncation of correlations,
which is clearly inconsistent with the observed divergence of the
correlation length $\xi$ and relaxation time $\tau$ as criticality is
approached: $\xi \sim |\lambda-\lambda_c|^{-\nu_\perp}$ and $\tau \sim
|\lambda-\lambda_c|^{-\nu_{||}}$, where $\nu_\perp>0$ and $\nu_{||}>0$
are critical exponents~\cite{Marro99}. In our simulations, correlation
lengths and relaxation times at criticality are limited only by system
size. At criticality also the survival probability ${\cal P}(t)$
decays as a power law (as opposed to an exponential fall in the
subcritical regime)~\cite{Marro99}, which can lead to long-lived
excitation waves, as illustrated in the single run of
Fig.~\ref{fig:singlerun}. We note that correlations among cortical
neurons several synapses distant from one another have been
experimentally observed~\cite{Roelfsema97} and associated with the
propagation of electrical waves~\cite{Ermentrout01}.

\section{\label{conclusion}Concluding remarks}

We have presented an analysis (with simulations and analytical
results) of the response function of the stochastic SIRS model on
hypercubic lattices. We confirmed that, as argued in
Ref.~\cite{Kinouchi06a}, the maximum dynamic range is obtained
precisely at the nonequilibrium phase transition where self-sustained
activity becomes stable. Moreover, since the response function at
criticality is governed by the critical exponent $\delta_h^{-1}$,
which for the DP universality class increases with $d$, the maximum
dynamic range obtained at a given dimension is a decreasing function
of $d$. We therefore corroborate the claim that networks with spatial
organization may have larger dynamic ranges than the random networks
for which these ideas were first developed~\cite{Kinouchi06a}.

This suggests the usefulness of low-dimensional arrays of excitable
units for artificial sensor design, as well as raises speculations
regarding the effective dimensionality of living neural networks. If
one admits that large dynamic ranges could be favored by natural
selection, organisms would tend to have their brains tuned at
criticality, and in this case the mystery of how Stevens' exponents
$<1$ arise would be solved: they would just be the critical exponent
$\delta_h^{-1}$~\cite{Kinouchi06a}.

In this context, our theoretical results join a recent flow of
experimental evidence which is compatible with neurons collectively
operating in a critical
regime~\cite{Deans02,Beggs03,Beggs04,Plenz07,Chialvo04,Eguiluz05}. It
is interesting to note that these experiments very often reveal
exponents close to mean-field values: for instance, in cultures and
acute slices of rat cortex, spontaneous activity occurs in avalanches
whose size distribution decays as a power law with an exponent
$3/2$~\cite{Beggs03,Beggs04,Plenz07}, which is the mean field result
for branching processes. Also, the response function of retinal
ganglion cells is well fitted by $\rho\sim
h^{0.58}$~\cite{Deans02,Furtado06}, yielding an exponent which is
remarkably close to the Stevens' exponent for light
intensity~\cite{Stevens} and the DP $\delta_h^{-1}$ exponent for
$d=4$. It would be interesting to investigate whether a small-world
connectivity could conciliate these results, on the one hand
preserving the local order observed in real neural networks
(see~\cite{Sporns04a} for a recent review) while on the other hand
allowing for mean-field exponents owing to a small density of
long-range connections.

\begin{acknowledgments}
V.R.V.A and M.C. acknowledge financial support from CAPES, Conselho
Nacional de Desenvolvimento Cient\'{\i}fico e Tecnol\'ogico (CNPq),
FACEPE, and special program PRONEX. The authors are grateful to
R. Dickman, M. A. F. Gomes, J. Joo, S. C. Ferreira Jr., O. Kinouchi,
D. R. Chialvo, and an anonymous referee for discussions and
suggestions.
\end{acknowledgments}

\appendix*

\section{Equations for the two-site approximation}

As mentioned previously, the equations in Ref.~\cite{Joo04b} can be
easily extended by including an external field $h$ as follows: 

\begin{eqnarray}
\label{evolucao_temporal_dos_pares_1}
\dot P_t(S_x,I_y) & = & \gamma P_t(R_x,I_y) - (\lambda + 1 + h)P_t(S_x,I_y) \nonumber \\
&& + hP_t(S_x,S_y) + \sum_{w \in {\cal N}^x(y)} \lambda P_t(S_x,S_y,I_w) \nonumber \\
&& - \sum_{w \in {\cal N}^y(x)} \lambda P_t(I_w,S_x,I_y) 
\end{eqnarray}
\begin{eqnarray}
\label{evolucao_temporal_dos_pares_2}
\dot P_t(S_x,R_y) & = & P_t(S_x,I_y) + \gamma P_t(R_x,R_y) \nonumber\\
&& - (\gamma + h) P_t(S_x,R_y)	\nonumber \\
&& - \sum_{w \in {\cal N}^y(x)}\lambda P_t(I_w,S_x,R_y) 
\end{eqnarray}
\begin{eqnarray}
\label{evolucao_temporal_dos_pares_3}
\dot P_t(R_x,I_y) & = & -(\gamma + 1) P_t(R_x,I_y) +  P_t(I_x,I_y)\nonumber \\
&& + hP_t(R_x,S_y) \nonumber \\
&& + \sum_{w \in {\cal N}^x(y)} \lambda P_t(R_x,S_y,I_w),
\end{eqnarray}
where ${\cal N}^x(y)$ is the neighborhood of $y$, excluding $x$. Note
that we can omit the equation for $P_t(I_x,I_y)$ because of the
normalization condition $\sum_{A} P_t(A_x,B_y)= P_t(B_y)$ [the
same reasoning applies to $P_t(S_x,S_y)$ and $P_t(R_x,R_y)$]. By
applying the two-site approximation to Eqs.~(\ref{eq:p0})-(\ref{eq:p2})
and
Eqs.~(\ref{evolucao_temporal_dos_pares_1})-(\ref{evolucao_temporal_dos_pares_3})
under homogeneity and isotropy assumptions~\cite{Joo04b}, as well as
the normalization condition $\sum_{A} P_t(A_x)=1$, we obtain

\begin{eqnarray}
\label{eq:par1}
\dot P_t(S) & = & \gamma - (\gamma + h)P_t(S) - \gamma P_t(I) \nonumber \\
&& - z \lambda P_t(S,I), \\
\label{eq:par2}
\dot P_t(I) & = & h P_t(S) - P_t(I) + z \lambda P_t(S,I), \\
\label{eq:par3}
\dot P_t(S,I) & = & hP_t(S) - (\lambda + 1 + 2h)P_t(S,I) -hP_t(S,R) \nonumber \\
&& + \gamma P_t(R,I) \nonumber \\
&& + (z-1)\lambda \frac {P_t(S,I)} {P_t(S)} \left[ P_t(S) - 2P_t(S,I) \right. \nonumber \\
&& \left. - P_t(S,R) \right], \\
\label{eq:par4}
\dot P_t(S,R) & = & \gamma - \gamma P_t(S) - \gamma P_t(I) + P_t(S,I) \nonumber \\
&& - (2 \gamma + h)P_t(S,R) - \gamma P_t(R,I) \nonumber \\
&& - (z - 1) \lambda \frac {P_t(S,I)P_t(S,R)} {P_t(S)}, 
\end{eqnarray}
\begin{eqnarray}
\label{eq:par5}
\dot P_t(R,I) & = & P_t(I) - P_t(S,I) + hP_t(S,R) - (2 + \gamma)P_t(R,I) \nonumber \\
&& + (z - 1) \lambda \frac {P_t(S,I)P_t(S,R)} {P_t(S)}.
\end{eqnarray}

The above equations form a closed system of ordinary differential
equations. In its fixed point, $\rho = \lim_{t\to\infty}P_t(I)$ is
shown to satisfy the cubic equation

\begin{equation}
\label{eq:cubic}
A_1 \rho^3 + A_2 \rho^2 + A_3 \rho + A_4 = 0\; ,
\end{equation}
where 

\begin{eqnarray}
	\label{A_1}
	A_1 & = & \gamma^2 \{\gamma^3 [z^2 (z-1) \lambda - z] + \gamma^2 [z (2 z^2-2 z-1) \lambda - 2 z \nonumber \\
               && - 1] + \gamma [2 z (z^2-z-1) \lambda -2 z - 1] + z [(z^2-z-1) \lambda \nonumber \\
               && - 1]\} - h \Big\{ h^2 (\gamma + 1)^4 + h [\gamma^5 z + \gamma^4 (z \lambda + 4 z + 3) + \nonumber \\
               && \gamma^3 (4 z \lambda + 7 z + 9) + \gamma^2 (6 z \lambda + 7 z + 9) + \gamma (4 z \lambda + \nonumber \\
               && 4 z + 3) + z \lambda + z] + \gamma \{\gamma^4 (z^2 \lambda + 2 z) + \gamma^3 [z \lambda (3 z + \nonumber \\
               && 2) + 6 z + 3] + 2 \gamma^2 [z (2 z+3)\lambda + 4z + 3] + 3 \gamma [z (z + \nonumber \\
               && 2) \lambda + 2 z + 1] + z [(z+2) \lambda +
               2)]\}\Big\}\; ,
\end{eqnarray}

\begin{eqnarray}
	\label{A_2}
	A_2 & = & \gamma \bigg\{z \gamma^2 \{\gamma^2 [-2 z (z-1)\lambda + z + 1] + \gamma [-(3 z^2 - 4 z \nonumber \\
               && - 1)\lambda + z + 3] - (2 z^2-3 z-1)\lambda + z + 1\} + \nonumber \\
               && h \Big\{3 h^2 (\gamma + 1)^3 + h [3 \gamma^4 z  + \gamma^3 (3 z \lambda + 11 z +6) + \nonumber \\
               && \gamma^2 (9 z \lambda + 16 z + 12) + \gamma (9 z \lambda + 11 z + 6) + 3 z \lambda + \nonumber \\
               && 3 z] + \gamma \{\gamma^3 [3 z^2 \lambda + z (z+4)] + \gamma^2 [4 z (2 z+1) \lambda + \nonumber \\
               && 2 z(z + 6) + 3] + \gamma [z (9 z+8) \lambda + 2 z(z + 6) + 3] + \nonumber \\
               && z [4 (z+1) \lambda + z + 4]\}\Big\}\bigg\}, 
\end{eqnarray}

\begin{eqnarray}
	\label{A_3}
	A_3 & = & \gamma^2 \bigg\{z^2 \gamma^2 \{\gamma [(z-1) \lambda - 1] + (z-2) \lambda - 1\} \nonumber \\
               && - h \Big\{3 h^2 (\gamma + 1)^2 + h[3 \gamma^3 z + \gamma^2 (3 z \lambda + 10 z + 3) + \nonumber \\
               && \gamma (6 z \lambda + 10 z + 3) + 3 z \lambda+3 z] + z\gamma\{\gamma^2 [3 z \lambda + \nonumber \\
               && 2 (z + 1)] + \gamma [(7 z+2) \lambda + 3 (z + 2)] + (5 z+2) \lambda + \nonumber \\
               && 2 (z + 1)\}\Big\}\bigg\}\; , 
\end{eqnarray}

\begin{eqnarray}
	\label{A_4}
	A_4 & = & h \gamma^3 \{h^2 (\gamma + 1) + h z [\gamma^2 + \gamma (\lambda + 3) + \lambda + 1] + \nonumber \\
               && z^2 \gamma [\gamma (\lambda + 1) + 2 \lambda + 1]\}.
\end{eqnarray}
Cardan's formula yields the solution of eq.~(\ref{eq:cubic}), from which
the dynamic range can be numerically obtained.


\bibliography{copelli}

\begin{thebibliography}{39}
\expandafter\ifx\csname natexlab\endcsname\relax\def\natexlab#1{#1}\fi
\expandafter\ifx\csname bibnamefont\endcsname\relax
  \def\bibnamefont#1{#1}\fi
\expandafter\ifx\csname bibfnamefont\endcsname\relax
  \def\bibfnamefont#1{#1}\fi
\expandafter\ifx\csname citenamefont\endcsname\relax
  \def\citenamefont#1{#1}\fi
\expandafter\ifx\csname url\endcsname\relax
  \def\url#1{\texttt{#1}}\fi
\expandafter\ifx\csname urlprefix\endcsname\relax\def\urlprefix{URL }\fi
\providecommand{\bibinfo}[2]{#2}
\providecommand{\eprint}[2][]{\url{#2}}

\bibitem[{\citenamefont{Hodgkin and Huxley}(1952)}]{HH52}
\bibinfo{author}{\bibfnamefont{A.~L.} \bibnamefont{Hodgkin}} \bibnamefont{and}
  \bibinfo{author}{\bibfnamefont{A.~F.} \bibnamefont{Huxley}},
  \bibinfo{journal}{J. Neurophysiol.} \textbf{\bibinfo{volume}{117}},
  \bibinfo{pages}{500} (\bibinfo{year}{1952}).

\bibitem[{\citenamefont{Stevens}(1975)}]{Stevens}
\bibinfo{author}{\bibfnamefont{S.~S.} \bibnamefont{Stevens}},
  \emph{\bibinfo{title}{Psychophysics: Introduction to its Perceptual, Neural
  and Social Prospects}} (\bibinfo{publisher}{Wiley, New York},
  \bibinfo{year}{1975}).

\bibitem[{\citenamefont{Rospars et~al.}(2000)\citenamefont{Rospars, L\'ansk\'y,
  Duchamp-Viret, and Duchamp}}]{Rospars00}
\bibinfo{author}{\bibfnamefont{J.-P.} \bibnamefont{Rospars}},
  \bibinfo{author}{\bibfnamefont{P.}~\bibnamefont{L\'ansk\'y}},
  \bibinfo{author}{\bibfnamefont{P.}~\bibnamefont{Duchamp-Viret}},
  \bibnamefont{and} \bibinfo{author}{\bibfnamefont{A.}~\bibnamefont{Duchamp}},
  \bibinfo{journal}{BioSystems} \textbf{\bibinfo{volume}{58}},
  \bibinfo{pages}{133} (\bibinfo{year}{2000}).

\bibitem[{\citenamefont{Rospars et~al.}(2003)\citenamefont{Rospars, L\'ansk\'y,
  Duchamp-Viret, and Duchamp}}]{Rospars03}
\bibinfo{author}{\bibfnamefont{J.-P.} \bibnamefont{Rospars}},
  \bibinfo{author}{\bibfnamefont{P.}~\bibnamefont{L\'ansk\'y}},
  \bibinfo{author}{\bibfnamefont{P.}~\bibnamefont{Duchamp-Viret}},
  \bibnamefont{and} \bibinfo{author}{\bibfnamefont{A.}~\bibnamefont{Duchamp}},
  \bibinfo{journal}{Eur. J. Neurosci.} \textbf{\bibinfo{volume}{18}},
  \bibinfo{pages}{1135} (\bibinfo{year}{2003}).

\bibitem[{\citenamefont{Cleland and Linster}(1999)}]{Cleland99}
\bibinfo{author}{\bibfnamefont{T.~A.} \bibnamefont{Cleland}} \bibnamefont{and}
  \bibinfo{author}{\bibfnamefont{C.}~\bibnamefont{Linster}},
  \bibinfo{journal}{Neural Computation} \textbf{\bibinfo{volume}{11}},
  \bibinfo{pages}{1673} (\bibinfo{year}{1999}).

\bibitem[{\citenamefont{Normann and Werblin}(1974)}]{Normann74}
\bibinfo{author}{\bibfnamefont{R.~A.} \bibnamefont{Normann}} \bibnamefont{and}
  \bibinfo{author}{\bibfnamefont{F.~S.} \bibnamefont{Werblin}},
  \bibinfo{journal}{J. Gen. Physiol.} \textbf{\bibinfo{volume}{63}},
  \bibinfo{pages}{37} (\bibinfo{year}{1974}).

\bibitem[{\citenamefont{Werblin}(1974)}]{Werblin74a}
\bibinfo{author}{\bibfnamefont{F.~S.} \bibnamefont{Werblin}},
  \bibinfo{journal}{J. Gen. Physiol.} \textbf{\bibinfo{volume}{63}},
  \bibinfo{pages}{62} (\bibinfo{year}{1974}).

\bibitem[{\citenamefont{Werblin and Copenhagen}(1974)}]{Werblin74b}
\bibinfo{author}{\bibfnamefont{F.~S.} \bibnamefont{Werblin}} \bibnamefont{and}
  \bibinfo{author}{\bibfnamefont{D.~R.} \bibnamefont{Copenhagen}},
  \bibinfo{journal}{J. Gen. Physiol.} \textbf{\bibinfo{volume}{63}},
  \bibinfo{pages}{88} (\bibinfo{year}{1974}).

\bibitem[{\citenamefont{Kim and Rieke}(2003)}]{Kim03}
\bibinfo{author}{\bibfnamefont{K.~J.} \bibnamefont{Kim}} \bibnamefont{and}
  \bibinfo{author}{\bibfnamefont{F.}~\bibnamefont{Rieke}}, \bibinfo{journal}{J.
  Neurosci.} \textbf{\bibinfo{volume}{23}}, \bibinfo{pages}{1506}
  (\bibinfo{year}{2003}).

\bibitem[{\citenamefont{Borst et~al.}(2005)\citenamefont{Borst, Flanagin, and
  Sompolinsky}}]{Borst05}
\bibinfo{author}{\bibfnamefont{A.}~\bibnamefont{Borst}},
  \bibinfo{author}{\bibfnamefont{V.~L.} \bibnamefont{Flanagin}},
  \bibnamefont{and}
  \bibinfo{author}{\bibfnamefont{H.}~\bibnamefont{Sompolinsky}},
  \bibinfo{journal}{Proc. Natl. Acad. Sci. {USA}}
  \textbf{\bibinfo{volume}{102}}, \bibinfo{pages}{6172} (\bibinfo{year}{2005}).

\bibitem[{\citenamefont{Deans et~al.}(2002)\citenamefont{Deans, Volgyi,
  Goodenough, Bloomfield, and Paul}}]{Deans02}
\bibinfo{author}{\bibfnamefont{M.~R.} \bibnamefont{Deans}},
  \bibinfo{author}{\bibfnamefont{B.}~\bibnamefont{Volgyi}},
  \bibinfo{author}{\bibfnamefont{D.~A.} \bibnamefont{Goodenough}},
  \bibinfo{author}{\bibfnamefont{S.~A.} \bibnamefont{Bloomfield}},
  \bibnamefont{and} \bibinfo{author}{\bibfnamefont{D.~L.} \bibnamefont{Paul}},
  \bibinfo{journal}{Neuron} \textbf{\bibinfo{volume}{36}}, \bibinfo{pages}{703}
  (\bibinfo{year}{2002}).

\bibitem[{\citenamefont{Furtado and Copelli}(2006)}]{Furtado06}
\bibinfo{author}{\bibfnamefont{L.~S.} \bibnamefont{Furtado}} \bibnamefont{and}
  \bibinfo{author}{\bibfnamefont{M.}~\bibnamefont{Copelli}},
  \bibinfo{journal}{Phys. Rev. E} \textbf{\bibinfo{volume}{73}},
  \bibinfo{pages}{011907} (\bibinfo{year}{2006}).

\bibitem[{\citenamefont{Copelli et~al.}(2002)\citenamefont{Copelli, Roque,
  Oliveira, and Kinouchi}}]{Copelli02}
\bibinfo{author}{\bibfnamefont{M.}~\bibnamefont{Copelli}},
  \bibinfo{author}{\bibfnamefont{A.~C.} \bibnamefont{Roque}},
  \bibinfo{author}{\bibfnamefont{R.~F.} \bibnamefont{Oliveira}},
  \bibnamefont{and} \bibinfo{author}{\bibfnamefont{O.}~\bibnamefont{Kinouchi}},
  \bibinfo{journal}{Phys. Rev. E} \textbf{\bibinfo{volume}{65}},
  \bibinfo{pages}{060901} (\bibinfo{year}{2002}).

\bibitem[{\citenamefont{Copelli et~al.}(2005)\citenamefont{Copelli, Oliveira,
  Roque, and Kinouchi}}]{Copelli05a}
\bibinfo{author}{\bibfnamefont{M.}~\bibnamefont{Copelli}},
  \bibinfo{author}{\bibfnamefont{R.~F.} \bibnamefont{Oliveira}},
  \bibinfo{author}{\bibfnamefont{A.~C.} \bibnamefont{Roque}}, \bibnamefont{and}
  \bibinfo{author}{\bibfnamefont{O.}~\bibnamefont{Kinouchi}},
  \bibinfo{journal}{Neurocomputing} \textbf{\bibinfo{volume}{65-66}},
  \bibinfo{pages}{691} (\bibinfo{year}{2005}).

\bibitem[{\citenamefont{Copelli and Kinouchi}(2005)}]{Copelli05b}
\bibinfo{author}{\bibfnamefont{M.}~\bibnamefont{Copelli}} \bibnamefont{and}
  \bibinfo{author}{\bibfnamefont{O.}~\bibnamefont{Kinouchi}},
  \bibinfo{journal}{Physica A} \textbf{\bibinfo{volume}{349}},
  \bibinfo{pages}{431} (\bibinfo{year}{2005}).

\bibitem[{\citenamefont{Kinouchi and Copelli}(2006)}]{Kinouchi06a}
\bibinfo{author}{\bibfnamefont{O.}~\bibnamefont{Kinouchi}} \bibnamefont{and}
  \bibinfo{author}{\bibfnamefont{M.}~\bibnamefont{Copelli}},
  \bibinfo{journal}{Nature Phys.} \textbf{\bibinfo{volume}{2}},
  \bibinfo{pages}{348} (\bibinfo{year}{2006}).

\bibitem[{\citenamefont{Wu et~al.}(2007)\citenamefont{Wu, Xu, and Wang}}]{Wu07}
\bibinfo{author}{\bibfnamefont{A.-C.} \bibnamefont{Wu}},
  \bibinfo{author}{\bibfnamefont{X.-J.} \bibnamefont{Xu}}, \bibnamefont{and}
  \bibinfo{author}{\bibfnamefont{Y.-H.} \bibnamefont{Wang}},
  \bibinfo{journal}{Phys. Rev. E} \textbf{\bibinfo{volume}{75}},
  \bibinfo{pages}{032901} (\bibinfo{year}{2007}).

\bibitem[{\citenamefont{Copelli and Campos}(2007)}]{Copelli07}
\bibinfo{author}{\bibfnamefont{M.}~\bibnamefont{Copelli}} \bibnamefont{and}
  \bibinfo{author}{\bibfnamefont{P.~R.~A.} \bibnamefont{Campos}},
  \bibinfo{journal}{Eur. Phys. J. B} \textbf{\bibinfo{volume}{56}},
  \bibinfo{pages}{273} (\bibinfo{year}{2007}).

\bibitem[{\citenamefont{Brecht et~al.}(2004)\citenamefont{Brecht, Schneider,
  Sakmann, and Margrie}}]{Brecht04}
\bibinfo{author}{\bibfnamefont{M.}~\bibnamefont{Brecht}},
  \bibinfo{author}{\bibfnamefont{M.}~\bibnamefont{Schneider}},
  \bibinfo{author}{\bibfnamefont{B.}~\bibnamefont{Sakmann}}, \bibnamefont{and}
  \bibinfo{author}{\bibfnamefont{T.~W.} \bibnamefont{Margrie}},
  \bibinfo{journal}{Nature} \textbf{\bibinfo{volume}{427}},
  \bibinfo{pages}{704} (\bibinfo{year}{2004}).

\bibitem[{\citenamefont{Hestrin and Galarreta}(2005)}]{Hestrin05}
\bibinfo{author}{\bibfnamefont{S.}~\bibnamefont{Hestrin}} \bibnamefont{and}
  \bibinfo{author}{\bibfnamefont{M.}~\bibnamefont{Galarreta}},
  \bibinfo{journal}{Trends Neurosci.} \textbf{\bibinfo{volume}{28}},
  \bibinfo{pages}{304} (\bibinfo{year}{2005}).

\bibitem[{\citenamefont{Friedrich and Korsching}(1997)}]{Friedrich97}
\bibinfo{author}{\bibfnamefont{R.~W.} \bibnamefont{Friedrich}}
  \bibnamefont{and} \bibinfo{author}{\bibfnamefont{S.~I.}
  \bibnamefont{Korsching}}, \bibinfo{journal}{Neuron}
  \textbf{\bibinfo{volume}{18}}, \bibinfo{pages}{737} (\bibinfo{year}{1997}).

\bibitem[{\citenamefont{Wachowiak and Cohen}(2001)}]{Wachowiak01}
\bibinfo{author}{\bibfnamefont{M.}~\bibnamefont{Wachowiak}} \bibnamefont{and}
  \bibinfo{author}{\bibfnamefont{L.~B.} \bibnamefont{Cohen}},
  \bibinfo{journal}{Neuron} \textbf{\bibinfo{volume}{32}}, \bibinfo{pages}{723}
  (\bibinfo{year}{2001}).

\bibitem[{\citenamefont{Joo and Lebowitz}(2004)}]{Joo04b}
\bibinfo{author}{\bibfnamefont{J.}~\bibnamefont{Joo}} \bibnamefont{and}
  \bibinfo{author}{\bibfnamefont{J.~L.} \bibnamefont{Lebowitz}},
  \bibinfo{journal}{Phys. Rev. E} \textbf{\bibinfo{volume}{70}},
  \bibinfo{pages}{036114} (\bibinfo{year}{2004}).

\bibitem[{\citenamefont{Marro and Dickman}(1999)}]{Marro99}
\bibinfo{author}{\bibfnamefont{J.}~\bibnamefont{Marro}} \bibnamefont{and}
  \bibinfo{author}{\bibfnamefont{R.}~\bibnamefont{Dickman}},
  \emph{\bibinfo{title}{Nonequilibrium Phase Transition in Lattice Models}}
  (\bibinfo{publisher}{Cambridge University Press},
  \bibinfo{address}{Cambridge}, \bibinfo{year}{1999}).

\bibitem[{\citenamefont{Adler and Duarte}(1987)}]{Adler&Duarte1987}
\bibinfo{author}{\bibfnamefont{J.}~\bibnamefont{Adler}} \bibnamefont{and}
  \bibinfo{author}{\bibfnamefont{J.~A. M.~S.} \bibnamefont{Duarte}},
  \bibinfo{journal}{Phys. Rev. B} \textbf{\bibinfo{volume}{35}},
  \bibinfo{pages}{7046} (\bibinfo{year}{1987}).

\bibitem[{\citenamefont{Adler et~al.}(1988)\citenamefont{Adler, Berger, Duarte,
  and Meir}}]{Adler&Duarte1988}
\bibinfo{author}{\bibfnamefont{J.}~\bibnamefont{Adler}},
  \bibinfo{author}{\bibfnamefont{J.}~\bibnamefont{Berger}},
  \bibinfo{author}{\bibfnamefont{J.~A. M.~S.} \bibnamefont{Duarte}},
  \bibnamefont{and} \bibinfo{author}{\bibfnamefont{Y.}~\bibnamefont{Meir}},
  \bibinfo{journal}{Phys. Rev. B} \textbf{\bibinfo{volume}{37}},
  \bibinfo{pages}{7529} (\bibinfo{year}{1988}).

\bibitem[{\citenamefont{Obukhov}(1980)}]{Obukhov1980}
\bibinfo{author}{\bibfnamefont{S.~P.} \bibnamefont{Obukhov}},
  \bibinfo{journal}{Physica A} \textbf{\bibinfo{volume}{101}},
  \bibinfo{pages}{145} (\bibinfo{year}{1980}).

\bibitem[{\citenamefont{Janssen}(1981)}]{Janssen81}
\bibinfo{author}{\bibfnamefont{H.~K.} \bibnamefont{Janssen}},
  \bibinfo{journal}{Z. Phys. B} \textbf{\bibinfo{volume}{42}},
  \bibinfo{pages}{151} (\bibinfo{year}{1981}).

\bibitem[{\citenamefont{Janssen}(1985)}]{Janssen85}
\bibinfo{author}{\bibfnamefont{H.~K.} \bibnamefont{Janssen}},
  \bibinfo{journal}{Z. Phys. B} \textbf{\bibinfo{volume}{58}},
  \bibinfo{pages}{311} (\bibinfo{year}{1985}).

\bibitem[{\citenamefont{Grassberger}(1982)}]{Grassberger82}
\bibinfo{author}{\bibfnamefont{P.}~\bibnamefont{Grassberger}},
  \bibinfo{journal}{Z. Phys. B} \textbf{\bibinfo{volume}{47}},
  \bibinfo{pages}{365} (\bibinfo{year}{1982}).

\bibitem[{\citenamefont{Greenberg and Hastings}(1978)}]{Greenberg78}
\bibinfo{author}{\bibfnamefont{J.~M.} \bibnamefont{Greenberg}}
  \bibnamefont{and} \bibinfo{author}{\bibfnamefont{S.~P.}
  \bibnamefont{Hastings}}, \bibinfo{journal}{{SIAM} J. Appl. Math.}
  \textbf{\bibinfo{volume}{34}}, \bibinfo{pages}{515} (\bibinfo{year}{1978}).

\bibitem[{\citenamefont{Roelfsema et~al.}(1997)\citenamefont{Roelfsema, Engel,
  K{\" o}nig, and Singer}}]{Roelfsema97}
\bibinfo{author}{\bibfnamefont{P.~R.} \bibnamefont{Roelfsema}},
  \bibinfo{author}{\bibfnamefont{A.~K.} \bibnamefont{Engel}},
  \bibinfo{author}{\bibfnamefont{P.}~\bibnamefont{K{\" o}nig}},
  \bibnamefont{and} \bibinfo{author}{\bibfnamefont{W.}~\bibnamefont{Singer}},
  \bibinfo{journal}{Nature} \textbf{\bibinfo{volume}{385}},
  \bibinfo{pages}{157} (\bibinfo{year}{1997}).

\bibitem[{\citenamefont{Ermentrout and Kleinfeld}(2001)}]{Ermentrout01}
\bibinfo{author}{\bibfnamefont{G.~B.} \bibnamefont{Ermentrout}}
  \bibnamefont{and}
  \bibinfo{author}{\bibfnamefont{D.}~\bibnamefont{Kleinfeld}},
  \bibinfo{journal}{Neuron} \textbf{\bibinfo{volume}{29}}, \bibinfo{pages}{33}
  (\bibinfo{year}{2001}).

\bibitem[{\citenamefont{Beggs and Plenz}(2003)}]{Beggs03}
\bibinfo{author}{\bibfnamefont{J.~M.} \bibnamefont{Beggs}} \bibnamefont{and}
  \bibinfo{author}{\bibfnamefont{D.}~\bibnamefont{Plenz}}, \bibinfo{journal}{J.
  Neurosci.} \textbf{\bibinfo{volume}{23}}, \bibinfo{pages}{11167}
  (\bibinfo{year}{2003}).

\bibitem[{\citenamefont{Beggs and Plenz}(2004)}]{Beggs04}
\bibinfo{author}{\bibfnamefont{J.~M.} \bibnamefont{Beggs}} \bibnamefont{and}
  \bibinfo{author}{\bibfnamefont{D.}~\bibnamefont{Plenz}}, \bibinfo{journal}{J.
  Neurosci.} \textbf{\bibinfo{volume}{24}}, \bibinfo{pages}{5216}
  (\bibinfo{year}{2004}).

\bibitem[{\citenamefont{Plenz and Thiagarajan}(2007)}]{Plenz07}
\bibinfo{author}{\bibfnamefont{D.}~\bibnamefont{Plenz}} \bibnamefont{and}
  \bibinfo{author}{\bibfnamefont{T.~C.} \bibnamefont{Thiagarajan}},
  \bibinfo{journal}{Trends Neurosci.} \textbf{\bibinfo{volume}{30}},
  \bibinfo{pages}{101} (\bibinfo{year}{2007}).

\bibitem[{\citenamefont{Chialvo}(2004)}]{Chialvo04}
\bibinfo{author}{\bibfnamefont{D.~R.} \bibnamefont{Chialvo}},
  \bibinfo{journal}{Physica A} \textbf{\bibinfo{volume}{340}},
  \bibinfo{pages}{756} (\bibinfo{year}{2004}).

\bibitem[{\citenamefont{Egu\'{\i}luz et~al.}(2005)\citenamefont{Egu\'{\i}luz,
  Chialvo, Cecchi, Baliki, and Apkarian}}]{Eguiluz05}
\bibinfo{author}{\bibfnamefont{V.~M.} \bibnamefont{Egu\'{\i}luz}},
  \bibinfo{author}{\bibfnamefont{D.~R.} \bibnamefont{Chialvo}},
  \bibinfo{author}{\bibfnamefont{G.~A.} \bibnamefont{Cecchi}},
  \bibinfo{author}{\bibfnamefont{M.}~\bibnamefont{Baliki}}, \bibnamefont{and}
  \bibinfo{author}{\bibfnamefont{A.~V.} \bibnamefont{Apkarian}},
  \bibinfo{journal}{Phys. Rev. Lett.} \textbf{\bibinfo{volume}{94}},
  \bibinfo{pages}{018102} (\bibinfo{year}{2005}).

\bibitem[{\citenamefont{Sporns et~al.}(2004)\citenamefont{Sporns, Chialvo,
  Kaiser, and Hilgetag}}]{Sporns04a}
\bibinfo{author}{\bibfnamefont{O.}~\bibnamefont{Sporns}},
  \bibinfo{author}{\bibfnamefont{D.~R.} \bibnamefont{Chialvo}},
  \bibinfo{author}{\bibfnamefont{M.}~\bibnamefont{Kaiser}}, \bibnamefont{and}
  \bibinfo{author}{\bibfnamefont{C.~C.} \bibnamefont{Hilgetag}},
  \bibinfo{journal}{Trends Cog. Sci.} \textbf{\bibinfo{volume}{8}},
  \bibinfo{pages}{418} (\bibinfo{year}{2004}).

\end{thebibliography}

\end{document}